\documentclass[showpacs,pra]{revtex4}

\usepackage{graphicx}
\usepackage{dcolumn}
\usepackage{bm}
\newcommand{\beq}{\begin{equation}}
\newcommand{\eeq}{\end{equation}}
\newcommand{\bea}{\begin{eqnarray}}
\newcommand{\eea}{\end{eqnarray}}
\newcommand{\bec}{\begin{center}}
\newcommand{\enc}{\end{center}}
\newcommand{\bfr}{\begin{flushright}}
\newcommand{\efr}{\end{flushright}}

\newcommand{\om}{\omega}

\newcommand{\Om}{\Omega}

\newcommand{\rmd}{{\rm d}}
\newcommand{\rmi}{{\rm i}}

\newcommand{\gr}{\gamma_{\rm r}}
\newcommand{\gd}{\gamma_{\rm d}}
\newcommand{\gn}{\gamma_{\rm n}}
\newcommand{\lssr}{\langle ss \rangle}
\newcommand{\sds}{\langle s^{\dagger}s \rangle}
\newcommand{\sdsp}{\langle s^{\dagger}s' \rangle}
\newcommand{\sdss}{\langle s^{\dagger}ss \rangle}
\newcommand{\sdssp}{\langle s^{\dagger}ss' \rangle}

\begin{document}
\title{Drastic effects of damping mechanisms 
on the third-order optical nonlinearity}
\author{Kazuki Koshino}
 \email{ikuzak@aria.mp.es.osaka-u.ac.jp}
\author{Hajime Ishihara}
\affiliation{
CREST, Japan Science and Technology Agency, 
4-1-8 Honcho, Kawaguchi, Saitama 332-0012, Japan\\
Department of Materials Engineering Science,
Osaka University, Toyonaka, Osaka 560-8531, Japan
}
\date{\today}
\begin{abstract}
We have investigated the optical response of superradiant atoms,
which undergoes three different damping mechanisms:
radiative dissipation ($\gr$),
dephasing ($\gd$), and nonradiative dissipation ($\gn$).
Whereas the roles of $\gd$ and $\gn$ are equivalent 
in the linear susceptibility $\chi^{(1)}$,
the third-order nonlinear susceptibility $\chi^{(3)}$
drastically depends on the ratio of $\gd$ and $\gn$:
When $\gd \ll \gn$, $\chi^{(3)}$ is essentially that of a single atom.
Contrarily, in the opposite case of $\gd \gg \gn$, 
$\chi^{(3)}$ suffers the size-enhancement effect
and becomes proportional to the system size.
\end{abstract}
\maketitle

There has been much interest in optical responses of finite-sized systems.
When the eigenstates of the system are delocalized in space,
the transition dipole moment between the ground state 
and the lowest excited state scales as $N^{1/2}$,
where $N$ is a parameter representing the system size~\cite{Henry}.
This $N^{1/2}$ scaling rule is the origin of 
unique optical responses of finite-sized systems, such as 
size-enhancement of third-order optical susceptibility, 
$\chi^{(3)}$~\cite{Hana,Taka,MukaPRA,MukaPRL,Ishi1,Ishi2}.
In conventional theories on the size-effects of optical responses,
whereas the size-dependences on eigenenergies 
and transition dipole moments were carefully taken into account,
damping effects were often treated rather crudely 
by simply introducing phenomenological damping constants.
However, because the system size would affect 
not only eigenenergies and transition dipole moments but also damping rates,
more rigorous treatment on the damping effects are desired.
Particularly, when the material is irradiated by resonant light fields,
the magnitude of the optical nonlinearity is 
strongly sensitive to the damping rates.
It is therefore indispensable to exclude phenomenology 
on the damping effects for quantitative evaluation of optical nonlinearity.

Here, targeting solid-state nonlinear optical devises in mind, 
we investigate the optical response of finite sized systems 
under a situation where the system suffers three different 
damping mechanisms: radiative dissipation, dephasing, and 
nonradiative dissipation. 
The latter two damping mechanisms are brought about 
by coupling to environmental degrees of freedom such as phonons.  
The effects of nonradiative dampings have been considered 
in detail in Ref.~\cite{Ishi1}, 
but the damping constants are introduced by hand, 
independently of the model. 
(The radiative damping is incorporated through 
the self-consistent Maxwell fields in their formalism~\cite{Ishi2}.) 
Spano {\it et. al.} pioneered the theories without phenomenology 
on the damping effects, where the damping dynamics of the system 
is explicitly defined in the model~\cite{MukaPRA,MukaPRL}. 
As for nonradiative damping effects, 
they introduced the homogeneous dephasing alone and 
no dissipation was explicitly treated. 
Although particular aspects of damping effects on 
the size-dependence of nonlinearity have been considered, 
the interplay of different damping mechanisms 
is still a subject of importance. 
To reveal the overall effects brought about by three different 
damping mechanisms, it is indispensable to treat the radiative 
and nonradiative dampings on equal footing and 
to make a clear distinction between dephasing and nonradiative dissipation. 
It is shown here that the third-order nonlinear response is 
drastically dependent on the ratio of two kinds of nonradiative 
damping rates: the ratio determines whether the size-enhancement 
of nonlinear response occurs or not. 
This fact implies that particular damping conditions could provide 
a novel resource for size-enhancement of nonlinear response.

The objective of this study is 
to investigate the third-order nonlinear optical response,
taking account of different damping mechanisms explicitly.
As a simplest model of a nonlinear optical system with finite size,
we consider a superradiant system composed by $N$ identical two-level systems
(hereafter referred to as ^^ ^^ atoms'')
with transition frequency $\Omega$~\cite{Dicke},
which suffers, individually at each atom, 
both dephasing and nonradiative dissipation.
Such damping mechanisms become particularly significant 
if the atoms are embedded in a solid-state environment,
{\it e. g.}, quantum dots in a microcavity.
The equation of motion for the density matrix $\rho$ of atoms
is the superradiant master equation~\cite{Haro,MukaPRA,MukaPRL}
supplemented with 
the terms describing dephasing and 
nonradiative dissipation~\cite{QO}.
It is given, omitting $\hbar$ and $\mu$ (transition dipole moment) 
for notational simplicity, by
\bea
\frac{\rmd \rho}{\rmd t} &=&
-\rmi[{\cal H}_0 + {\cal H}_{\rm int}(t),\rho]
+({\cal L}_{\rm r}+{\cal L}_{\rm d}+{\cal L}_{\rm n})\rho, \label{eq:eqm1}
\\
{\cal H}_0 &=& \sum_j \Omega s_j^{\dagger} s_j,
\\
{\cal H}_{\rm int}(t) &=& E(t)S^{\dagger} + E^{\ast}(t)S,
\label{eq:Hint}
\\
{\cal L}_{\rm r}\rho &=&
\frac{\gr}{2}(2S\rho S^{\dagger} -S^{\dagger}S\rho  - \rho S^{\dagger}S),
\label{eq:Lr}
\\
{\cal L}_{\rm d}\rho &=&
-\frac{\gd}{2}\sum_j[s_j^{\dagger} s_j, [s_j^{\dagger} s_j, \rho]],
\label{eq:Ld}
\\
{\cal L}_{\rm n}\rho &=&
\frac{\gn}{2}\sum_j(2s_j \rho s_j^{\dagger} 
-s_j^{\dagger}s_j \rho  - \rho s_j^{\dagger}s_j),
\label{eq:Ln}
\eea
where $s_j^{\dagger}$ and $s_j$ are 
the Pauli creation and annihilation operators at $j$th atom,
and $S=\sum_j s_j$ is the collective operator.
$E(t)$ represents the positive frequency part of the applied electric field,
and the rotating wave approximation is used in ${\cal H}_{\rm int}(t)$.
$\gr$, $\gd$ and $\gn$ represent the single-atom rates of radiative decay,
dephasing, and nonradiative dissipation, respectively. 
It is of note that 
the atoms interact with the electromagnetic field 
via the collective operator [see Eqs.~(\ref{eq:Hint}) and (\ref{eq:Lr})]
whereas dephasing and nonradiative dissipations 
occurs independently in each atom
[see Eqs.~(\ref{eq:Ld}) and (\ref{eq:Ln})].

Based on this model, we investigate the linear 
and the third-order nonlinear optical responses.
To the end of investigating up to third-order response,
we are concerned with the following expectation values:
$\langle s_i \rangle$,
$\langle s_i s_j \rangle$,
$\langle s_i^{\dagger} s_j \rangle$, and
$\langle s_i^{\dagger} s_j s_k \rangle$,
where the expectation value of an operator $A$
is given by $\langle A \rangle = {\rm Tr}\{\rho A\}$.
Remembering the fact that all atoms are equivalent,
the number of independent variables are greatly reduced.
For example, it is apparent that $\langle s_i \rangle$ 
is independent of the site index $i$.
We use the following notations:
\bea
\langle s_i \rangle &=& \langle s \rangle,
\\
\langle s_i s_j \rangle &=&
\left\{\matrix{0 & (i=j)  \cr
\langle ss \rangle & (i \neq j)}\right.,
\\
\langle s_i^{\dagger} s_j \rangle &=&
\left\{\matrix{\langle s^{\dagger}s \rangle & (i=j) \cr
\langle s^{\dagger}s' \rangle & (i \neq j)}\right.,
\\
\langle s_i^{\dagger} s_j s_k \rangle &=&
\left\{\matrix{0 & (j=k) \cr
\langle s^{\dagger}ss \rangle & (j\neq k, i=j \ {\rm or} \ k) \cr
\langle s^{\dagger}ss' \rangle & (i\neq j, j\neq k, k\neq i) 
}\right..
\eea
The equation of motion for $\langle A \rangle$ is given,
using Eq.~(\ref{eq:eqm1}), by
$\frac{\rmd}{\rmd t}\langle A \rangle =
-\rmi \langle [A, {\cal H}_0+{\cal H}_{\rm int}(t)] \rangle
+(\gr/2)\langle [S_+,A]S_- + S_+[A,S_-] \rangle
-(\gd/2)\sum_j \langle [[A,s_j^{\dagger} s_j],s_j^{\dagger} s_j] \rangle
+(\gn/2)\sum_j \langle [s_j^{\dagger},A]s_j + 
s_j^{\dagger}[A,s_j] \rangle$.
Defining $\Gamma_{a,b,c}$ by
\beq
\Gamma_{a,b,c} = (a\gr+b\gd+c\gn)/2, 
\eeq
the equations of motion for $\langle s \rangle$ {\it etc}
are given as follows:
\begin{widetext}
\bea
d \langle s_1 \rangle/dt &=& 
(-\rmi \Omega -\Gamma_{N,1,1})\langle s_1 \rangle 
-\rmi E, 
\\
d \langle ss \rangle/dt &=& 
(-2\rmi \Omega -\Gamma_{2N-2,2,2})\langle ss \rangle
-2 \rmi E \langle s_1 \rangle,
\\
d \langle s^{\dagger}s \rangle/dt &=& 
(\rmi E^{\ast} \langle s_1 \rangle + c.c.)
-\Gamma_{2,0,2}\langle s^{\dagger}s \rangle
-\Gamma_{2N-2,0,0}\langle s^{\dagger}s' \rangle,
\label{eq:eqm_sds}
\\
d \langle s^{\dagger}s' \rangle/dt &=& 
(\rmi E^{\ast} \langle s_1 \rangle + c.c.)
-\Gamma_{2,0,0}\langle s^{\dagger}s \rangle
-\Gamma_{2N-2,2,2}\langle s^{\dagger}s' \rangle,
\label{eq:eqm_sdsp}
\\
d \langle s_3 \rangle/dt &=& 
(-\rmi \Omega -\Gamma_{N,1,1})\langle s_3 \rangle 
+2\rmi E \langle s^{\dagger}s \rangle 
+ \Gamma_{2N-2,0,0}\langle s^{\dagger}ss \rangle,
\\
d \langle s^{\dagger}ss \rangle/dt &=& 
(-\rmi \Omega -\Gamma_{N+2,1,3})\langle s^{\dagger}ss \rangle
-\Gamma_{2N-4,0,0}\langle s^{\dagger}ss' \rangle
+ \rmi E^{\ast} \langle ss \rangle
- \rmi E (\langle s^{\dagger}s \rangle 
+ \langle s^{\dagger}s' \rangle),
\\
d \langle s^{\dagger}ss' \rangle/dt &=& 
(-\rmi \Omega -\Gamma_{3N-6,3,3})\langle s^{\dagger}ss' \rangle
-\Gamma_{4,0,0}\langle s^{\dagger}ss \rangle
+ \rmi E^{\ast} \langle ss \rangle
- 2\rmi E \langle s^{\dagger}s' \rangle,
\eea
\end{widetext}
where $\langle s_1 \rangle$ and $\langle s_3 \rangle$ denotes 
the first- and third-order components of $\langle s \rangle$.
Although not explicitly indicated,
$\langle ss \rangle$, $\langle s^{\dagger}s \rangle$ and $\langle s^{\dagger}s' \rangle$
($\langle s^{\dagger}ss \rangle$ and $\langle s^{\dagger}ss' \rangle$)
in the above equations are the second- (third-)order quantities.
It is of note that,
in the above equations of motion,
the dependence on the system size $N$
appears only through the enhancement of $\gr$.

We can easily obtain the stationary solutions 
of these simultaneous equations.
Assuming that $E(t)$ is monochromatic as $E(t) \sim e^{-\rmi \om t}$, 
and introducing $f_{a,b,c}(\om)$ by
\beq
f_{a,b,c}(\om)=( \om-\Omega+\rmi\Gamma_{a,b,c})^{-1},
\eeq
$\langle s_1 \rangle$, $\langle ss \rangle$, $\langle s^{\dagger}s \rangle$ 
and $\langle s^{\dagger}s' \rangle$ are given as follows:
\begin{widetext}
\bea
\langle s_1 \rangle &=& f_{N,1,1}(\om) E,
\\
\langle ss \rangle &=& f_{N,1,1}(\om)f_{N-1,1,1}(\om) E^2,
\\
\left( \matrix{\langle s^{\dagger}s \rangle 
\cr \langle s^{\dagger}s' \rangle} \right)
&=&
\frac{N\gr+\gd+\gn}{\gr(\gd+N\gn)+\gn(\gd+\gn)} 
\left(\matrix{\gd+\gn \cr \gn} \right)|f_{N,1,1}(\om)|^2|E|^2.
\label{eq:sds}
\eea
Eq.~(\ref{eq:sds}) shows that,
in determining time-independent quantities such as $\sds$ and $\sdsp$,
the ratio of damping constants play crucial roles.
This feature was also observed in conventional theories 
on the nonlinear susceptibilities 
using phenomenological damping constants~\cite{Hana,Taka,Ishi1}.
In terms of the second-order quantities,
$\langle s^{\dagger}ss \rangle$ and $\langle s^{\dagger}ss' \rangle$
are given by
\beq
\left( \matrix{\langle s^{\dagger}ss \rangle 
\cr \langle s^{\dagger}ss' \rangle} \right)
=
\left(\matrix{
f^{-1}_{N+2,1,3}(\om) & \rmi \Gamma_{2N-4,0,0} \cr
\rmi \Gamma_{4,0,0} & f^{-1}_{3N-6,3,3}(\om)
}\right)^{-1}
\left(\matrix{-E^{\ast}\langle ss \rangle 
+ E \langle s^{\dagger}s \rangle
+ E \langle s^{\dagger}s' \rangle
\cr 
-E^{\ast} \langle ss \rangle
+ 2 E \langle s^{\dagger}s' \rangle}\right),
\label{eq:sdss}
\eeq
and $\langle s_3 \rangle$ is given, in terms of 
$\langle s^{\dagger}s \rangle$ and 
$\langle s^{\dagger}ss \rangle$, by
\beq
\langle s_3 \rangle = f_{N,1,1}(\om) 
[-2E \langle s^{\dagger}s \rangle 
+\rmi \Gamma_{2N-2,0,0} \langle s^{\dagger}ss \rangle].
\eeq
Thus, we obtain the linear and third-order susceptibilities 
per one atom as follows:
\bea
\chi^{(1)}(\om) &=& \frac{\langle s_1 \rangle}{E}=f_{N,1,1}(\om),
\label{eq:chi1}
\\
\chi^{(3)}(\om) &=& \frac{\langle s_3 \rangle}{|E|^2E}
=f_{N,1,1}(\om) 
\left[-2\frac{\langle s^{\dagger}s \rangle}{|E|^2}  
+\rmi \Gamma_{2N-2,0,0} \frac{\langle s^{\dagger}ss \rangle}{|E|^2E}\right],
\label{eq:chi3}
\eea
\end{widetext}
both of which are free from 
phenomenological treatment on the damping effects. 
In the following part of this study,
we discuss how $\chi^{(3)}$ depends on the relaxation parameters
($\gr$, $\gd$, $\gn$) and the number $N$ of atoms.
Recent nanotechnologies aim to fabricate clean quantum systems
with long coherence times,
and extensive efforts on suppressing $\gd$ and $\gn$ are being made.
In the following part of this study, we restrict our attention to a case 
where $\gd$ and $\gn$ are well suppressed 
and satisfy $(\gd, \gn)\ll\gr$.

Firstly, we discuss the limiting case of $\gd \to 0$.
In this limit, it is easily confirmed that
$\langle s^{\dagger}s \rangle = \langle s^{\dagger}s' \rangle
(=|f_{N,0,1}(\om)|^2 |E|^2)$
and 
$\langle s^{\dagger}ss \rangle = \langle s^{\dagger}ss' \rangle
(=|f_{N,0,1}(\om)|^2 f_{N-1,0,1}(\om) |E|^2E)$.
These equalities imply that $N$ atoms respond to 
the electric field cooperatively, as a spin $N/2$ object.
$\chi^{(3)}$ is reduced to the following form:
\beq
\chi^{(3)}_{\gd \to 0}(\om)=
-2|f_{N,0,1}(\om)|^2 f_{N,0,1}(\om)
f_{N-1,0,1}(\om) f^{-1}_{0,0,1}(\om).
\label{eq:gd0}
\eeq
This equation reveals that 
$\chi^{(3)}_{\gd \to 0}$ depends on the system size $N$
only through the enhancement of $\gr$.
In the off-resonant frequency regions,
$\chi^{(3)}_{\gd \to 0} \simeq -2/(\om -\Omega)^3$,
which is independent of $N$.
Thus, in the limit of $\gd \to 0$,
the optical nonlinearity is essentially that of a single atom,
except for minor corrections around the resonant frequency region.

\begin{figure}
\includegraphics{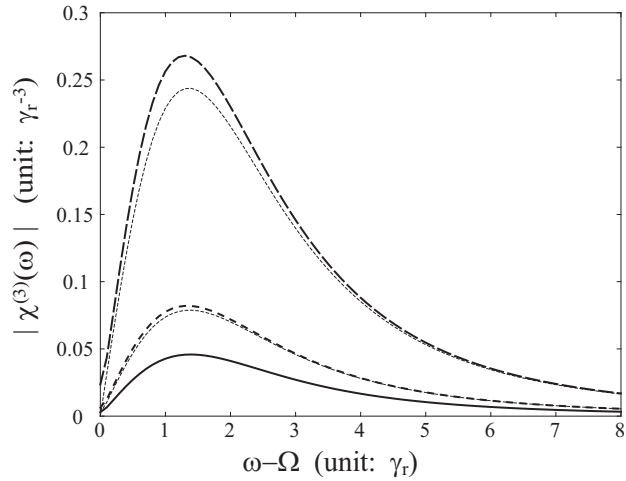}
\caption{\label{fig:chi3}
Comparison of $|\chi^{(3)}|$ and its approximate value 
$|\tilde{\chi}^{(3)}|$, when $N=5$.
The solid, dotted and broken lines show $|\chi^{(3)}|$
for ($\gd/\gr$, $\gn/\gr$)=(0, 0.1), (0.05, 0.05) and (0.1, 0).
The thin dotted lines show $|\tilde{\chi}^{(3)}|$ with the same parameters.
Note that $\tilde{\chi}^{(3)}$ coincides 
with $\chi^{(3)}$ for ($\gd/\gr$, $\gn/\gr$)=(0, 0.1).
}\end{figure}

Next, we consider a more general case of $\gd \neq 0$.
We should remark that, when $(\gd, \gn)\ll\gr$ is satisfied,
$\sds$ and $\sdsp$ are reduced to the following forms:
\bea
\sds & \simeq & \frac{N\gd + N\gn}{\gd + N\gn}
|f_{N,1,1}(\om)|^2 |E|^2
\\
\sdsp & \simeq & \frac{N\gn}{\gd + N\gn}
|f_{N,1,1}(\om)|^2 |E|^2
\eea
In case of $\gd \ll \gn$, all of 
$\lssr$, $\sds$, and $\sdsp$ are of the same order ($\sim |f|^2|E|^2$),
and the nonlinear susceptibility is given by Eq.~(\ref{eq:gd0}).
Contrarily, in the opposite case of $\gd \gg \gn$, 
$\sds$ becomes much larger than $\lssr$ and $\sdsp$.
($\sds \sim N|f|^2|E|^2$, whereas 
$\lssr$ and $\sdsp$ $\sim |f|^2|E|^2$.)
Then, Eqs. (\ref{eq:sdss}) and (\ref{eq:chi3}) suggest that 
$\sdss$, $\sdssp$ and $\chi^{(3)}$ 
become almost proportional to $\sds$.
Using the fact that $\sds$ is magnified by a factor 
$(N\gd + N\gn)/(\gd + N\gn)$ in comparison with the $\gd \to 0$ case,
we obtain the following approximate expression 
of the third-order nonlinear susceptibility:
\beq
\tilde{\chi}^{(3)}(\om) \simeq
\frac{N\gd + N\gn}{\gd + N\gn} \chi^{(3)}_{\gd \to 0}(\om).
\label{eq:app}
\eeq
In Fig.~\ref{fig:chi3}, 
the approximate susceptibility $\tilde{\chi}^{(3)}$ 
is compared with the rigorous susceptibility ${\chi}^{(3)}$.
The figure demonstrates that $\tilde{\chi}^{(3)}$
serves as a good approximation of ${\chi}^{(3)}$.

Now we discuss the implications of Eq.~(\ref{eq:app}).
As far as the linear optical response is questioned,
Eq.~(\ref{eq:chi1}) indicates that
the roles of dephasing ($\gd$) and nonradiative dissipation ($\gn$)
are equivalent in determining the linear optical response.
In contrast, when one questions the nonlinear optical response,
the roles of two damping mechanisms are no more equivalent:
The prefactor of the RHS of Eq.~(\ref{eq:app}) indicates that 
magnitude of $\chi^{(3)}$ is sensitive to the ratio $\gd/\gn$,
even if both $\gd$ and $\gn$ are much smaller than $\gr$.
When $\gd \ll \gn$, 
$\chi^{(3)}$ is essentially independent of the system size $N$.
Contrarily, when $\gd \gg \gn$, 
$\chi^{(3)}$ suffers the size-enhancement effect.
This observation demonstrates that it is indispensable 
for quantitative evaluation of nonlinear susceptibility
to discriminate two damping mechanisms
and to treat them non-phenomenologically.

\begin{figure}
\includegraphics{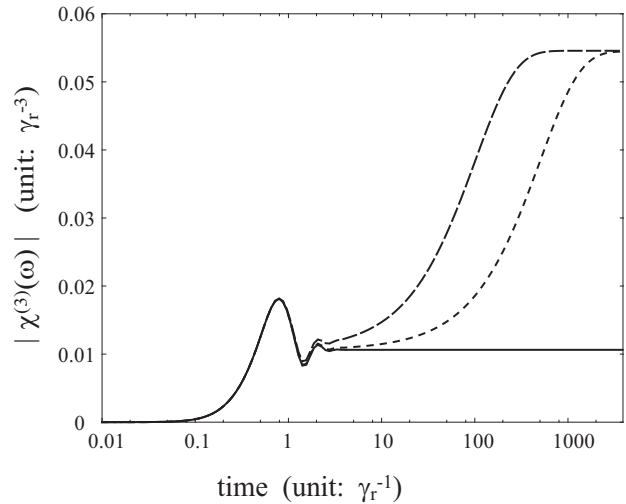}
\caption{\label{fig:trans}
Temporal behavior of $|\chi^{(3)}|$,
assuming that irradiation of monochromatic electric field, 
$E(t)=E e^{-\rmi\om t}$, starts at $t=0$.
The following parameters are used:
$N=5$, $\om = \Om+5\gr$, $\gn=0$, and $\gd/\gr=0$ (solid line),
0.01 (dotted line), and 0.05 (broken line).
Transition to the enhanced value takes place at $t \sim N/(2\gd)$.
}\end{figure}

One might feel uneasy about the fact 
that $\chi^{(3)}$ is indefinite at $\gd = \gn = 0$.
In order to resolve this problem,
we investigate the transient optical response 
by considering a situation where monochromatic field, 
$E(t)=E e^{-\rmi \om t}$, is switched on at $t=0$.
By inspecting Eqs.~(\ref{eq:eqm_sds}) and (\ref{eq:eqm_sdsp}),
we can find that $\chi^{(3)}$ has two relaxation rates,
$N \gr (=\tau_1^{-1})$ and $2\gd/N+2\gn (=\tau_2^{-1})$,
before attaining to its stationary value.
The temporal behavior of $|\chi^{(3)}|$ is 
numerically pursued in Fig.~\ref{fig:trans}.
The figure clarifies that 
$\chi^{(3)}$ first relaxes to the unenhanced value 
within a short time ($t \lesssim \tau_1$),
and the size-enhancement effect emerges gradually 
in the later stage ($t \sim \tau_2$).
When $\gd = \gn = 0$ (solid line in Fig.~\ref{fig:trans}), 
the size-enhancement does not take place, 
and $\chi^{(3)}=\chi^{(3)}_{\gd \to 0}$ forever.
More generally,
when one is concerned with a transient behavior ($t \lesssim \tau_2$),
the size-enhancement effect is not expected;
it is expected only after a long time, $t \gtrsim \tau_2$.

Finally, we comment on the relevance to previous studies.
Mathematically, $\chi^{(3)}$ should be evaluated
by the stationary solutions of equations of motion
for $\langle s_i \rangle$, $\langle s_i^{\dagger}s_j \rangle$, 
{\it etc}~\cite{MukaPRL}.
Because these quantities generally have dependence on the site index $i$,
it is usually difficult to obtain analytic expression of $\chi^{(3)}$.
(By taking the eigenstates of the system Hamiltonian 
${\cal H}_0$ as the basis,
one may diagonalize the unitary part of Eq.~(\ref{eq:eqm1}).
However, the basis generally does not diagonalize 
the damping part of Eq.~(\ref{eq:eqm1}) simultaneously~\cite{MukaPRA}.
The conventional expansion for $\chi^{(3)}$ \cite{Shen} is 
obtained by approximately neglecting the off-diagonal part.)
In the model of our study, 
an analytic form of $\chi^{(3)}$ is obtained 
without any approximation
by virtue of symmetry of the system,
and it is revealed that tiny difference in the damping rates
in Eqs.~(\ref{eq:eqm_sds}) and (\ref{eq:eqm_sdsp})
may result in drastically different optical response,
as observed in Fig.~\ref{fig:trans}.
A novel prediction in the present study is that
the size-enhancement may take place, 
even when there is no transfer of excitations among the atoms.

In summary, we have analyzed the third-order nonlinear susceptibility
of superradiant atoms,
which undergoes three different damping mechanisms:
radiative dissipation ($\gr$),
dephasing ($\gd$), and nonradiative dissipation ($\gn$).
The analysis is based on the superradiant master equation
supplemented with effects of dephasing and nonradiative dissipation
[Eqs.(\ref{eq:eqm1})-(\ref{eq:Ln})].
The linear susceptibility $\chi^{(1)}$ and 
the third-order susceptibility $\chi^{(3)}$
per one atom are given by Eqs.~(\ref{eq:chi1}) and (\ref{eq:chi3}),
and $\chi^{(3)}$ is well approximated by Eq.~(\ref{eq:app}).
Whereas the roles of $\gd$ and $\gn$ are equivalent 
in $\chi^{(1)}$ [see Eq.~(\ref{eq:chi1})],
they are no more equivalent in $\chi^{(3)}$ [see Eq.~(\ref{eq:app})]:
$\chi^{(3)}$ depends on the ratio $\gd/\gn$.
When $\gd \ll \gn$, $\chi^{(3)}$ is essentially that of a single atom.
Contrarily, when $\gd \gg \gn$, 
$\chi^{(3)}$ suffers the size-enhancement effect
and becomes proportional to the system size $N$.
These observations indicate that, 
for qualitative evaluation of $\chi^{(3)}$,
it is indispensable to distinguish $\gd$ and $\gn$ clearly,
and to handle them in a non-phenomenological manner.

The authors are grateful to K. Edamatsu
for fruitful discussions.
This research is partially supported by
Japan Society of Promotion of Science,
Grant-in-Aid for Scientific Research (A), 16204018, 2004.

\end{document}